  \documentclass[final,3p,times,twocolumn]{elsarticle}

\usepackage{amssymb}
\usepackage{amsmath,amssymb,exscale}
\usepackage{graphicx}
\usepackage{epsfig}
\usepackage{multicol}
\usepackage{color}
\usepackage{mathrsfs}
\usepackage{hyperref}
\hypersetup{colorlinks,bookmarksopen,bookmarksnumbered,citecolor=rossos,
linkcolor=black,pdfstartview=FitH,urlcolor=blus}
\usepackage{amsfonts}
\usepackage{slashed}
\usepackage{blindtext}
 \usepackage{fancyhdr}
\usepackage{hyperref}

\definecolor{blus}{cmyk}{1,1,0,0.6}
\definecolor{verdes}{cmyk}{0.99,0,0.59,0.82}
\definecolor{rossos}{cmyk}{0,1,1,0.55}
\definecolor{greeny}{cmyk}{0.99,0,0.59,0.98}

\def\bp{\bar M_{\rm Pl}}
\def\Lag{\mathscr{L}}

\def\be{\begin{equation}}
\def\ee{\end{equation}}
\def\bea{\begin{eqnarray}}
\def\eea{\end{eqnarray}}
\def\ba{\begin{array} }
\def\ea{\end{array}}
\def\bac{\begin{array} {c}}
\def\bacc{\begin{array} {cc}}
\def\baccc{\begin{array} {ccc}}
\def\x{\textcolor{red}}
\definecolor{red}{rgb}{1,0,0}

\newcommand{\mub}{\bar{\mu}}

\def\hhref#1{\href{http://arxiv.org/abs/#1}{arXiv:#1}} 

\journal{the arXiv}

\begin{document}

\begin{frontmatter}




\title{\vspace{-2cm}\huge Classical and Quantum  Initial Conditions for Higgs Inflation}


\author{\vspace{1cm}{\large Alberto Salvio}~$^{1}$ and {\large Anupam Mazumdar}~$^{2,~3}$}

\address{\normalsize \vspace{0.2cm}$^{1}$~Departamento de F\'isica Te\'orica, Universidad Aut\'onoma de Madrid\\ and Instituto de F\'isica Te\'orica IFT-UAM/CSIC,  Madrid, Spain.  \\
$^{2}$~Consortium for fundamental Physics, Lancaster University, Lancaster, LA1 4YB, UK\\
 $^{3}$~IPPP, Durham University, Durham, DH1 3LE, UK
 \\ 
 \vspace{0.3cm}
 {\it {\small Report number: IFT-UAM/CSIC-15-063}}
 \vspace{-0.7cm}
 }

\begin{abstract}
We investigate whether Higgs inflation  can occur in the Standard Model starting from  natural initial conditions or not.  The Higgs has a non-minimal coupling to the Ricci scalar. We confine our attention to the regime where quantum Einstein gravity effects are small in order to have results that are independent of the ultraviolet completion of gravity.  At the classical level we find 
no tuning  is required to have a successful Higgs inflation, provided the initial homogeneity condition is satisfied. On the other hand, at the quantum level we obtain that the renormalization  for large non-minimal coupling requires an additional degree of freedom that transforms Higgs inflation into Starobinsky $R^2$ inflation, unless a tuning of the initial values of the running parameters is made.
\end{abstract}


\end{frontmatter}


\section{Introduction}\label{introduction}

\vspace{-0.05cm}

Inflation~\cite{Guth:1980zm,Linde:1981mu, Albrecht:1982wi} is perhaps one of the most natural way to stretch the initial quantum vacuum fluctuations to the size of the current Hubble patch, seeding the initial perturbations for the cosmic microwave background (CMB) radiation and large scale structure in the universe~\cite{Ade:2015xua} (for a theoretical treatment, see~\cite{Mukhanov:1990me}). Since inflation dilutes all matter it is pertinent that after the end of inflation the universe is filled with the right thermal degrees of freedom, i.e. the Standard Model (SM) degrees of freedom (for a review on pre- and post-inflationary dynamics, see~\cite{Mazumdar:2010sa}). The most economical 
way to achieve this would be via the vacuum energy density stored within the SM Higgs, whose properties are now being measured
at the Large Hadron Collider (LHC)~\cite{Aad:2012tfa,Chatrchyan:2012ufa}. Naturally, the decay of the Higgs would create all the SM quarks and leptons observed within the visible sector of the universe. Albeit, with just alone SM Higgs and minimal coupling to gravity,  it is hard to explain the temperature anisotropy observed in the CMB radiation without invoking 
physics beyond the SM~\footnote{Within supersymmetry it is indeed possible to invoke the flat direction composed of the 
Higgses to realize inflation with minimal gravitational interaction, see~\cite{Chatterjee:2011qr}, which can explain the current CMB observations.}.

However, a very interesting possibility may arise  within the SM  if the Higgs were to couple to gravity non-minimally - such as in the 
context of extended inflation~\cite{La:1989za}, which has recently received particular attention after the Higgs discovery at the LHC in the context of 
Higgs inflation~\cite{Bezrukov:2007ep}.  By tuning this non-minimal coupling constant, $\xi$, between the Ricci scalar of the Einstein-Hilbert term 
and the SM Higgs, it is possible to explain sufficient amount of e-folds of inflation and also fit other observables such as 
the amplitude of temperature anisotropy and the spectral tilt in the CMB data. Indeed, this is very nice and satisfactory, except that the non-minimal 
coupling, $\xi$, turns out to be very large (at the classical level $\xi\sim 10^{4}$) in order to explain the CMB observables. This effectively 
redefines the Planck's constant during inflation, and invites new challenges for this model,
whose consequences have been debated vigorously in many papers, such as~\cite{crit}.

One particular consequence of such large non-minimal coupling is that there is a new scale in the theory, $\bar M_{\rm Pl}/\sqrt{\xi}$, lower than the standard reduced Planck mass, 
$\bar M_{\rm Pl}\approx 2.435\times 10^{18}$~GeV.
Typically inflation occurs above this scale, the Higgs field takes a
vacuum expectation value (VEV) above $\bar M_{\rm Pl}/\sqrt{\xi}$ in order to sustain inflation sufficiently. In fact, the inflaton potential, in the Einstein frame,
approaches a constant plateau for sufficiently large field values. Effectively, the inflaton becomes a flat direction, 
where it does not cost any energy for the field to take any VEV beyond this cut-off.

Given this constraint on the initial VEV of the inflaton and the new scale, we wish to address two particularly relevant issues concerning 
the Higgs inflation model~\cite{Bezrukov:2007ep}, one on the classical front and the other on the quantum front.

\vspace{0.2cm}

\noindent$\bf I.$
 Classically, a large VEV of the inflaton does not pose a big problem as long 
as the initial energy density stored in the inflaton system, in the Einstein frame, is below the cut-off of the theory. Since, the potential energy remains bounded 
 below this cut-off, the question remains - what should be the classical initial condition for the kinetic energy of the inflaton?

A-priori there is no reason for the inflaton to move slowly on the plateau, therefore the question we wish to settle in this paper is what should be the 
range of phase space allowed for a sustainable inflation to occur with almost a flat potential? The aim of this paper is to address this classical initial 
condition problem~\footnote{Some single monomial potentials and exponential potentials exhibit a classic example of late time attractor where the 
inflaton field approaches a slow roll phase from large initial kinetic energy, see~\cite{attractor,Kofman:2002cj}.}. Here we strictly assume homogeneity of the universe from the very beginning; we do not raise the issue of initial homogeneity condition required for a successful inflation; this issue has been  discussed earlier in a generic inflationary context in many classic papers (see~\cite{Linde:1985ub,Goldwirth:1990iq}). In our paper, instead we look into the possibility of initial phase space for a 
{\it spatially flat} universe, and study under what pre-inflationary conditions Higgs inflation could prevail.

\vspace{0.2cm}

\noindent$\bf II.$ At quantum level, the original Higgs model poses a completely different challenge. A large $\xi$ will inevitably modify 
the initial action. One may argue that there will be quantum corrections to the Ricci scalar, $R$,  such as a Higgs-loop 
correction - leading to a quadratic in curvature action, i.e. $ R + \alpha R^2$  type correction, where $\alpha$ is a constant, whose magnitude 
we shall discuss in this paper. The analysis is based on the renormalization group equations (RGEs) of the SM parameters and the gravitational interactions. By restricting for simplicity the study to operators up to dimension 4, the RGE analysis  will yield a gravitational action that will become very similar to the Starobinsky type inflationary model~\cite{Starobinsky:1980te}~\footnote{
In principle, large $\xi$ may  also yield higher derivative corrections up to quadratic in order, see~\cite{Biswas:2011ar}, and also higher curvature corrections, but in this paper, we will  consider for simplicity the lowest order corrections. We will argue that the $\alpha R^2$ is necessarily generated  unless one is at the critical point of Ref. \cite{Bezrukov:2014bra} or invokes a fine-tuning on the initial values of the running parameters.}.

 One of the features of theories with curvature squared terms is that there are extra degrees of freedom involved in the problem, besides the SM
ones and the graviton. There is another scalar mode arising from $R^2$, which will also participate during inflation. The question then arises when this new scalar degree of freedom becomes dominant dynamically, and play the role of an inflaton creating the initial density perturbations?

\vspace{0.2cm}

The aim of this paper will be to address both the classical and quantum issues.

We briefly begin our discussion with essential ingredients of Higgs inflation in section~\ref{model}, then we discuss the classical pre-inflationary initial conditions 
for Higgs inflation in section~\ref{pre-inflation}. In this section, we discuss both analytical~\ref{analytic}, and numerical results~\ref{numerics}. In 
section~\ref{quantum}, we discuss the quantum correction to the original Higgs inflation model, i.e. we discuss the RGEs of the Planck mass in subsection~\ref{planck-mass}, SM parameters in \ref{RGE-SM}, and the gravitational correction arising due to large $\xi$ in subsection~\ref{RGE}, respectively.
 We briefly discuss our results and consequences for inflation in subsection~\ref{Starobinsky}, before concluding our paper. 

\section{The model}\label{model}

Let us define the Higgs inflation model \cite{Bezrukov:2007ep}. The action is
\begin{equation} S= \int d^4x\sqrt{-g}\left[\Lag_{\rm SM}-\left(\frac{\bp^2 }{2}+\xi |\mathcal{H}|^2\right)R\right], \label{Jordan-frame-total}\end{equation}
where  $\Lag_{\rm SM}$ is the SM Lagrangian minimally coupled to gravity,  $\xi$ is the parameter that determines the non-minimal coupling between the Higgs and the Ricci scalar $R$, and $\mathcal{H}$ is the Higgs doublet. 
The part of the action that depends on the metric and the Higgs field  {\it only} (the scalar-tensor part) is
\begin{equation} S_{\rm st} = \int d^4x\sqrt{-g}\left[|\partial \mathcal{H}|^2-V-\left(\frac{\bp^2 }{2}+\xi |\mathcal{H}|^2\right)R\right], \label{Jordan-frame}\end{equation}
where  
 $V=\lambda (|\mathcal{H}|^2-v^2/2)^2$ is the Higgs potential and $v$ is the electroweak Higgs VEV.  We take a sizable non-minimal coupling, $\xi>1$, because this is required by inflation as we will see.

The non-minimal coupling $-\xi |\mathcal{H}|^2 R$ can be eliminated through the {\it conformal} transformation
\begin{equation} g_{\mu \nu}\rightarrow   \Omega^{-2}  g_{\mu \nu}, \quad \Omega^2= 1+\frac{2\xi |\mathcal{H}|^2}{\bp^2}. \label{transformation}\end{equation}
The original frame, where the Lagrangian has the form in (\ref{Jordan-frame-total}), is called the Jordan frame, while the one where gravity is canonically normalized (obtained with the transformation above) is called the Einstein frame.
In the unitary gauge, where the only scalar field is the radial mode $\phi \equiv \sqrt{2|\mathcal{H}|^2}$,  we have (after the conformal transformation)
\begin{equation} S_{\rm st} = \int d^4x\sqrt{-g}\left[K \frac{(\partial \phi)^2}{2}-\frac{V}{\Omega^4}-\frac{\bp^2 }{2}R\right], \end{equation}
where $K\equiv \left(\Omega^2+6\xi^2\phi^2/\bp^2\right)/\Omega^4$.
The non-canonical Higgs kinetic term can be made canonical through the (invertible) field redefinition $\phi=\phi(\chi)$ defined by
\begin{equation} \frac{d\chi}{d\phi}= \sqrt{\frac{\Omega^2+6\xi^2\phi^2/\bp^2}{\Omega^4}},\label{chi}\end{equation}
with the conventional condition $\phi(\chi=0)=0$. One can find a closed expression of $\chi$ as a function of $\phi$:
\bea \chi(\phi) &=&  \bp \sqrt{\frac{1+6 \xi }{\xi}} \text{sinh}^{-1}\left[\frac{\sqrt{\xi  (1+6 \xi )}\phi }{\bp}\right]\nonumber\\&&-\sqrt{6} \bp \text{tanh}^{-1}\left[\frac{\sqrt{6} \xi \phi }{\sqrt{\bp^2+\xi  (1+6 \xi )\phi ^2}}\right].\eea
Thus, $\chi$ feels a potential 
\begin{equation} U\equiv \frac{V}{\Omega^4}=\frac{\lambda(\phi(\chi)^2-v^2)^2}{4(1+\xi\phi(\chi)^2/\bp^2)^2}\label{U} .\end{equation}

Let us now recall how slow-roll inflation emerges. From (\ref{chi}) and (\ref{U}) it follows \cite{Bezrukov:2007ep} that $U$ is exponentially flat when $\chi \gg \bp$, which is the key property to have inflation. Indeed, for such high field values the slow-roll parameters 
\be \epsilon \equiv\frac{\bp^2}{2} \left(\frac{1}{U}\frac{dU}{d\chi}\right)^2, \quad \eta \equiv \frac{\bp^2}{U} \frac{d^2U}{d\chi^2}
\label{epsilon-def}\ee
are guaranteed to be small. Therefore, the region in field configurations where $\chi > \bp$ (or equivalently \cite{Bezrukov:2007ep} $\phi> \bp/\sqrt{\xi}$) corresponds to inflation. We will investigate whether successful sow-roll inflation emerges also for large initial field kinetic energy in the next section. Here we simply assume that the time derivatives are small.

All the parameters of the model can be fixed through experiments and observations, including $\xi$ \cite{Bezrukov:2007ep, Bezrukov:2008ut}. $\xi$ can be obtained by requiring that the measured power spectrum \cite{Ade:2015xua},
\begin{equation}P_{R}= \frac{U/ \epsilon}{24\pi^2 \bp^4}= (2.14 \pm 0.05) \times 10^{-9} , \label{normalization} \end{equation}
is reproduced for a field value $\phi=\phi_{\rm b}$ corresponding to an appropriate number of e-folds of inflation \cite{Bezrukov:2008ut}:
\begin{equation}N=\int_{\phi_{\rm end}}^{\phi_{\rm b}}\frac{U}{\bp^2}\left(\frac{dU}{d\phi}\right)^{-1}\left(\frac{d\chi}{d\phi}\right)^2d\phi\approx 59, \label{e-folds}\end{equation}
where $\phi_{\rm end}$ is the field value at the end of inflation, that is 
\be \epsilon(\phi_{\rm end}) \approx 1. \label{inflation-end}\ee
 For $N=59$, by using the classical potential we obtain
  \be \xi = (5.02\mp0.06)\times10^4 \sqrt{\lambda},\qquad (N=59)\label{xi large}\ee 
where the uncertainty corresponds to the experimental uncertainty in Eq. (\ref{normalization}). Note that $\xi$ depends on $N$:
 \bea \xi &=& (4.61\mp0.06)\times10^4 \sqrt{\lambda},\qquad (N=54) \\ \xi &=& (5.43\mp0.06)\times10^4 \sqrt{\lambda}.\qquad (N=64)\eea 
  This result indicates that $\xi$ has to be much larger than one because $\lambda \sim 0.1$ (for precise determinations of this coupling in the SM see Refs. \cite{Degrassietal,Buttazzo:2013uya}).



\section{Pre-inflationary dynamics: classical analysis}\label{pre-inflation}

Let us now analyze the dynamics of this classical system in the homogeneous case without making any assumption on the initial value of the time 
derivative $\dot \chi$. We will assume that the universe is sufficiently homogeneous to begin inflation.

In the Einstein frame $S_{\rm st}$  is given by:
\be S_{\rm st} =\int d^4x\sqrt{-g}\left[\frac{(\partial \chi)^2}{2}-U-\frac{\bp^2 }{2}R\right],   \ee
where $U$ is the Einstein frame potential given in Eq.  (\ref{U}).

 Let us assume a universe with three dimensional translational and rotational symmetry, that is a Friedmann-Robertson-Walker (FRW) metric
 \be ds^2 = dt^2 -a(t)^2 \left[\frac{dr^2}{1- k r^2} +r^2(d\theta^2 +\sin^2\theta d\varphi^2)\right],  \label{FRW}\ee
 with $k=0,\pm 1$. 

Then the Einstein equations  and the scalar equations imply the following equations for $a(t)$ and the spatially homogeneous field $\chi(t)$  
  \bea
  \ddot \chi  + 3H\dot \chi+U'  &=&0,\label{scalarEq} \\  
  \frac{\dot a^2+k}{a^2}-\frac{ \dot\chi^2+2U}{6 \bar M_{\rm Pl}^2} &=&0 \label{EE1}, \\ 
  \frac{k}{a^2}-\dot H-\frac{\dot \chi^2}{ 2\bar M_{\rm Pl}^2}&=&0. \label{EE2}\eea
where $H\equiv \dot a/a$, a dot denotes a derivative with respect to $t$ and a prime is a derivative with respect to $\chi$.
 Notice that  Eq. (\ref{scalarEq}) tells us that $\chi$ cannot be constant before inflation unless $U$ is flat. 
   From Eqs. (\ref{scalarEq}) and (\ref{EE1}) one can derive (\ref{EE2}), which is therefore dependent.

Thus, we have to solve the following system with initial conditions
\be \label{system} \bac  \left\{\bac \dot \Pi  + 3H\Pi+U'  = 0, \qquad\quad \Pi(\bar t) = \overline\Pi,  \\  \\ \hspace{-0.cm}  \dot\chi= \Pi, \qquad\qquad \qquad \qquad\,\, \chi(\bar t)=\overline \chi,\\  \\ \hspace{-0.3cm}\dot a^2+k=\frac{a^2}{6 \bar M_{\rm Pl}^2} ( \Pi^2+2U),\quad a(\bar t)=\bar a,
 \ea \right.  \ea \ee
where $\bar t$ is some initial time before inflation and $\overline \chi$, $\bar \Pi$ and $\bar a$ are the initial conditions for the three dynamical variables. In the case $k=0$ the previous system can be reduced to a single second order equation. Indeed, by setting $k=0$ in Eq. (\ref{EE1}) and inserting it in Eq. (\ref{scalarEq}), one obtains
\be \ddot\chi +\sqrt{\frac{3\dot\chi^2+6U}{2\bp^2}}\dot\chi +U'  = 0, \qquad (k=0).\label{eq-k=0}\ee
 This equation has to be solved with two initial conditions (for $\chi$ and $\dot\chi$). The initial condition for $a$ is not needed in this case as its overall normalization does not have a physical meaning for $k=0$.

We confine our attention to the regime where  quantum Einstein gravity  corrections are small: 
\be  U \ll \bp^4 , \quad \dot\chi^2\ll \bp^4,\quad \frac{|k|}{a^2}\ll \bp^2  \label{smallQG}\ee
such that we can ignore the details of the ultraviolet (UV) completion of Einstein gravity~\footnote{The conditions in (\ref{smallQG}) may  not be necessary in scenarios where gravitational interactions are softened at energies much below $\bp$  and remain small at and above $\bp$ \cite{Giudice:2014tma}. However, here we do not want to rely on any specific quantum gravity theory.}. However, we do not always require to be initially in a slow-roll regime. The first and second  conditions in (\ref{smallQG}) come from the requirement that the energy-momentum tensor is small (in units of the Planck scale) so that it does not source a large curvature; the third condition ensures that the three-dimensional curvature is also small. The first condition is automatically fulfilled by the Higgs inflation potential, Eq. (\ref{U}): the quartic coupling $\lambda$ is small \cite{Bezrukov:2012sa, Degrassietal,Buttazzo:2013uya} and the non-minimal coupling $\xi$ is large (see Eq. (\ref{xi large})).  The second and third conditions in (\ref{smallQG})   are implied by the requirement of starting from an (approximately) de Sitter space, which is maximally symmetric; therefore we do not consider them as a fine-tuning in the initial conditions. In de Sitter we have to set $k=0$ and $\dot  H=0$, which then implies $\dot \chi =0$ from Eq. (\ref{EE2}). Notice also that we cannot start from an exact de Sitter, given Eq. (\ref{scalarEq}):  the potential $U$ is almost, but not exactly flat in the large field case  (see Eq. (\ref{U})).

In order for the Higgs to trigger inflation sooner or later one should have a slow-roll regime, where the kinetic energy is small compared to the potential energy, $\dot\chi^2/2 \ll U$, and the field equations are approximately
\be  \frac{\dot a^2+k}{a^2} \approx \frac{U}{3  \bar M_{\rm Pl}^2} ,\quad  \dot \chi\approx -\frac{1}{3H}U', \quad \mbox{(slow-roll equations).} \label{slow-roll-eq}\ee  The conditions  for this to be true are
\be \dot\chi^2 \ll 2U, \quad  |\ddot\chi| \ll 3 |H \dot\chi|
\qquad \mbox{(slow-roll regime).}\label{slow-roll-con}\ee
We will use these conditions rather than the standard $\epsilon\ll 1$ and $\eta \ll1$ as we  do not assume a priori a small kinetic energy.


\subsection{Analytic approximations in simple cases}\label{analytic}

Let us assume, for simplicity,  that the parameter $k$ in the FRW metric vanishes, i.e. a spatially flat metric,
and consider the case  $\dot\chi^2 \gg U$, such that the potential energy can be neglected compared to the kinetic energy.
 In this case, combining Eqs. (\ref{EE1}) and (\ref{EE2}) gives 
\be  \dot H+3H^2+\frac{2k}{a^2}=0,\qquad (\dot\chi^2 \gg U),\ee
which for spatially flat curvature, $k=0$, leads to 
\be  H(t)= \frac{\bar H}{1+3\bar H(t-\bar{t})},\qquad (\dot\chi^2 \gg U, \,\, k=0),\label{HlargeK}\ee
where $\bar H\equiv H(\bar t)$. By inserting this result into Eq. (\ref{EE2}), we find 
\be \dot\chi^2= \frac{6\bp^2 {\bar H}^2}{\left[1+3\bar H(t-\bar{t})\right]^2},\qquad (\dot\chi^2 \gg U, \,\, k=0).\label{dotchiLargeK}\ee
that is the kinetic energy density scales as $1/t^2$ by taking into account the time dependence of $H$.  This result \cite{Linde:1985ub} tells us that an initial condition with large kinetic energy is attracted towards one with smaller kinetic energy, but it also shows that dropping the potential energy cannot be a good approximation for arbitrarily large times.
 Moreover, notice that Eqs. (\ref{HlargeK}) and (\ref{dotchiLargeK}) imply 
\be \ddot\chi = - 3H \dot\chi \ee
so the dynamics is {\it not} approaching the second condition in (\ref{slow-roll-con}).  Therefore, the argument above is not conclusive and  we need to solve the equations with $U$ included in order to see if the slow-roll regime is an attractor.



\begin{figure}[t]
 \vspace{0.3cm}
\begin{center}
 \includegraphics[scale=0.53]{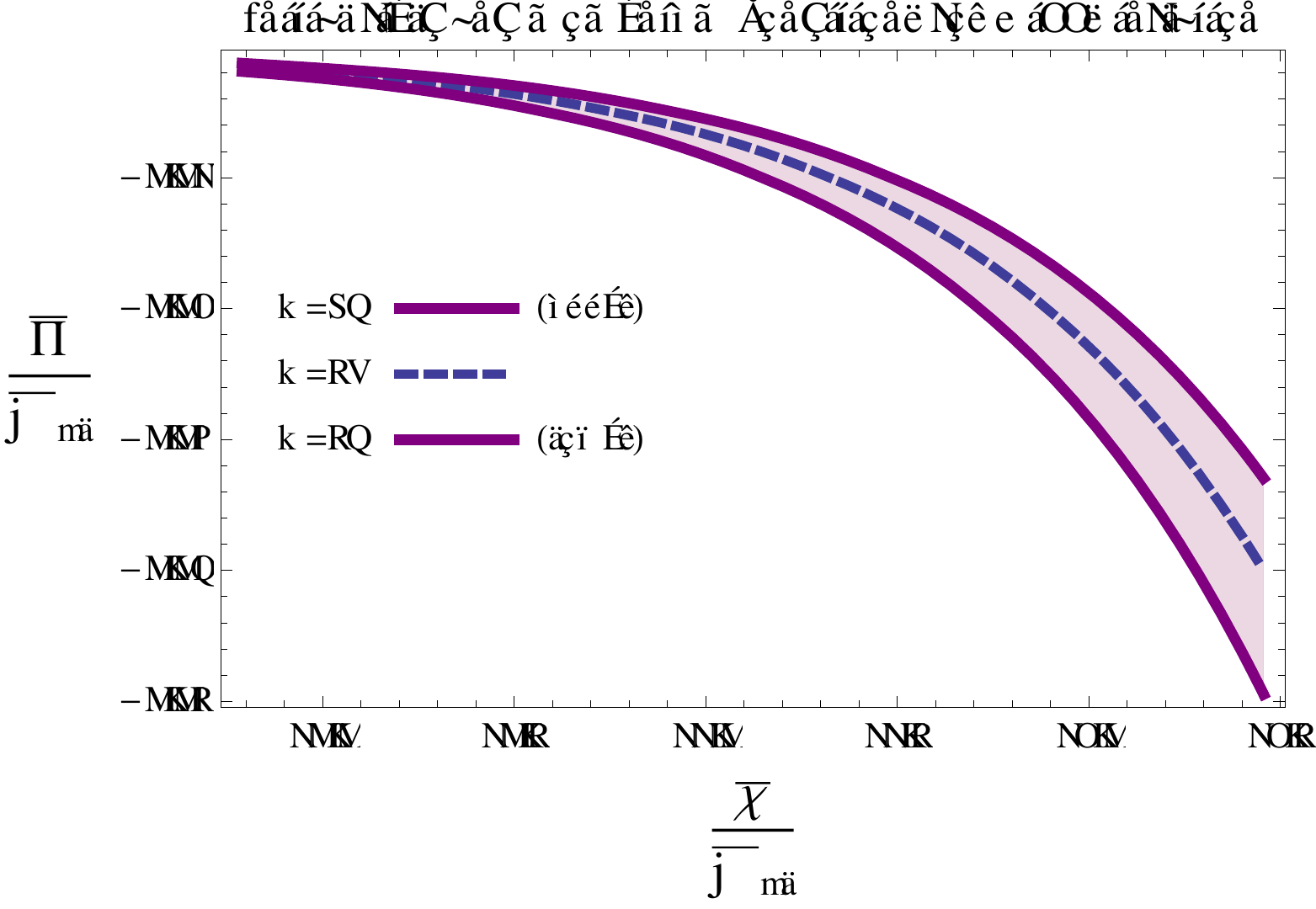}   
 \end{center}
 \vspace{-0.5cm}  
   \caption{\small Initial conditions $\overline \chi$ and $\overline \Pi$ for the Higgs field and its momentum $\Pi\equiv\dot\chi$ respectively. The thickness of the lines corresponds to $2\sigma$ uncertainty in the value of the power spectrum, Eq. (\ref{normalization}). }
\label{HIInitialCondition}
 \vspace{0.2cm}
\end{figure}

\subsection{Numerical studies}\label{numerics}

We studied numerically the system in (\ref{system})  assuming $k=0$; this case is realistic and is the simplest one: it does not require an initial condition for $a$. We found that even for an initial kinetic energy density $\overline\Pi^2$ of order $10^{-3}\bp^4$  (which we regard as the maximal  order of magnitude to have negligibly small quantum gravity), one should start from an initial field value $\overline\chi$ of order $10 \bp$ to inflate the universe for an appropriate number of e-folds, i.e. $N=59$.  This value of $\overline\chi$ is only one order of magnitude bigger than the one needed in the ordinary  case, $\overline\Pi^2 \ll U(\overline\chi) \sim10^{-10} \bp$, where the initial kinetic energy is much smaller than the potential energy.

Fig. \ref{HIInitialCondition} presents these results more quantitatively. There the initial  conditions for $\overline \Pi$  have been chosen to be negative because positive values favor slow-roll even with respect to the case where the initial kinetic energy is much smaller than the potential energy:  this is because the potential in Eq. (\ref{U}) is an increasing function of $\chi$ for $\chi\gg v$.

We conclude that at the classical level Higgs inflation does not suffer from a worrisome fine-tuning problem for the initial conditions.

\section{Quantum corrections}\label{quantum}

The theory in Eq. (\ref{Jordan-frame-total}) is not renormalizable. This means that quantum corrections $ \Delta\Gamma$ at a given order in perturbation theory can generate terms that are not combinations of those in the classical action $S$. In formulae the (quantum) effective action is given by:
\be \Gamma = S + \Delta\Gamma \ee 
where $S + \Delta\Gamma$ cannot generically be reproduced by substituting the  parameters in $S$  with some renormalized quantities.

A UV completion requires the existence of additional degrees of freedom that render the theory renormalizable or even finite. Much below the scale of this new physics, the effective action can be  
approximated by an expansion of the form 
\be \Delta\Gamma = \int d^4x \sqrt{-g}\left( \delta \Lag_2 + \delta \Lag_4 +\dots \right)   \ee
where $ \delta \Lag_n$ represents a combination of dimension $n$ operators.

We consider the one-loop corrections   
 generated by all fields of the theory, both the matter fields and gravity. Our purpose is to apply it to inflationary and pre-inflationary dynamics. We approximate  $\Delta\Gamma$ by including all operators up to dimension 4:  
 \be \Delta\Gamma \approx \int d^4x \sqrt{-g}\left( \delta \Lag_2 + \delta \Lag_4\right).   \ee
 This is the simplest approximation that allows us to include the dynamics of the Higgs field and possess scale invariance at high energies and high Higgs field values (up to running effects).
 We have 
 \bea \delta \Lag_2 &=&  -\frac{\delta \bp^2}{2} R \\  \delta \Lag_4 &=&\alpha  R^2 +\beta \left(\frac13 R^2 -R_{\mu\nu}R^{\mu\nu}\right)\nonumber \\ &&+  \delta Z_\mathcal{H} |\partial \mathcal{H}|^2 - \delta \lambda |\mathcal{H}|^4 - \delta \xi |\mathcal{H}|^2 R  + \dots \eea
 where for each parameter $p_c$ in the classical action we have introduced a corresponding quantum correction $\delta p$ and the dots represent the additional terms due to the fermions and gauge fields of the SM. Notice that we have added general~\footnote{$R_{\mu\nu\rho\sigma}R^{\mu\nu\rho\sigma}$ is a linear combination of $R^2$, $R_{\mu\nu}R^{\mu\nu}$ and a total derivative.} quantum corrections that are quadratic in the curvature tensors as they are also possible dimension 4 operators. These are parameterized by two dimensionless couplings $\alpha$ and $\beta$. We have neglected $v$ as it is very small compared to inflationary energies.
 
 Our purpose is now to determine the RGEs for the renormalized couplings $$p= p_c +\delta p$$ as well as for the new couplings  $\alpha$ and $\beta$ generated by quantum corrections. Indeed the RGEs encode the leading quantum corrections. We will use the dimensional regularization (DR) scheme  
 to regularize the loop integrals and the modified minimal subtraction ($\overline{\rm MS}$) scheme to renormalize away the divergences. This as usual leads to a renormalization scale that we denote with $\mub$.
 

  \subsection{RGE of the Planck mass}\label{planck-mass}
 In the absence of the dimensionful parameter $v$, the only possible contributions to the RGE of $\bp$ are the rainbow and the seagull  diagram contributions to the graviton propagator due to gravity itself: the rainbow topology is the one of Fig. \ref{HiggsContrToGraviton}, while the seagull one is obtained by making the two vertices of Fig. \ref{HiggsContrToGraviton} coincide without deforming the loop.

  The seagull diagram vanishes as it is given by combinations of loop integrals of the form 
 \be \int d^d k \frac{k_\mu k_\nu}{k^2 + i \epsilon}, \qquad \int d^d k \frac{1}{k^2 + i \epsilon}, \ee
 where $d$ is the space-time dimension in DR. These types of loop integrals vanish in DR. The rainbow diagram does not contribute to the RGE of $\bp$ either.  The reason is that each graviton propagator carries a factor of $1/\bp^2$ and each graviton vertex carries a factor of $\bp^2$  (because the graviton kinetic term $-\bp^2 R/2$ is proportional to $\bp^2$): the rainbow diagram has two graviton propagators and two vertices, therefore this contribution is dimensionless and cannot contribute to the RGE of a dimensionful quantity. We conclude that $\bp$ does not run in this case. This argument assumes that the graviton wave function renormalization is trivial, which we have checked to be the case at the one-loop level at hand.

 \subsection{RGEs of SM parameters}\label{RGE-SM}
 
 Having neglected $v$ all SM parameters are dimensionless and thus cannot receive contributions from loops involving graviton propagators (that carry a factor of $1/\bp^2$). Therefore, the SM RGEs apply and can be found (up to the three-loop level) in a convenient form in the appendix of Ref. \cite{Buttazzo:2013uya}.

 \subsection{RGEs of gravitational couplings}\label{RGE}
 
 Finally, we consider the RGEs for $\xi$, $\alpha$ and $\beta$. The one of $\xi$ does not receive contribution from loops involving graviton propagators as they carry a factor of $1/\bp^2$ and $\xi$ is dimensionless. So the RGE of $\xi$ receives contribution from the SM couplings and $\xi$ itself only \cite{Shapiro,Salvio:2014soa}:
  \be (4\pi)^2 \frac{d\xi}{d\ln\mub}=(1+6\xi)\left(y_t^2-\frac34 g_2^2 - \frac{3}{20} g_1^2+2\lambda\right),\ee
  where $y_t$ is the top Yukawa coupling and $g_3$, $g_2$ and $g_Y=\sqrt{3/5}g_1$ are the gauge couplings of SU(3)$_c$, SU(2)$_L$ and U(1)$_Y$ respectively. 
  
  The RGEs of $\alpha$ and $\beta$ receive two contributions: one from pure gravity loops (a rainbow and a seagull diagram), which we denote with $\beta^{g}$,  and one from matter loops, $\beta^m$: 
   \bea (4\pi)^2 \frac{d\alpha}{d\ln\mub} &=& \beta_{\alpha}^{g}+ \beta_{\alpha}^m,
\\
(4\pi)^2 \frac{d\beta}{d\ln\mub} &=& \beta_{\beta}^{g} + \beta_{\beta}^m.\eea
One finds  \cite{'tHooft:1974bx} 
\be \beta_{\alpha}^{g}= -\frac14,\qquad \beta_{\beta}^{g}=\frac7{10},\ee
and in the SM \cite{Salvio:2014soa}
\be   \beta_{\alpha}^m= - \frac{(1+6\xi)^2}{18},\qquad \beta_{\beta}^m= \frac{283}{60}.\label{beta-m}\ee

\begin{figure}[t]
\begin{center}
\vspace{-1cm}
 \includegraphics[scale=0.26, angle =90]{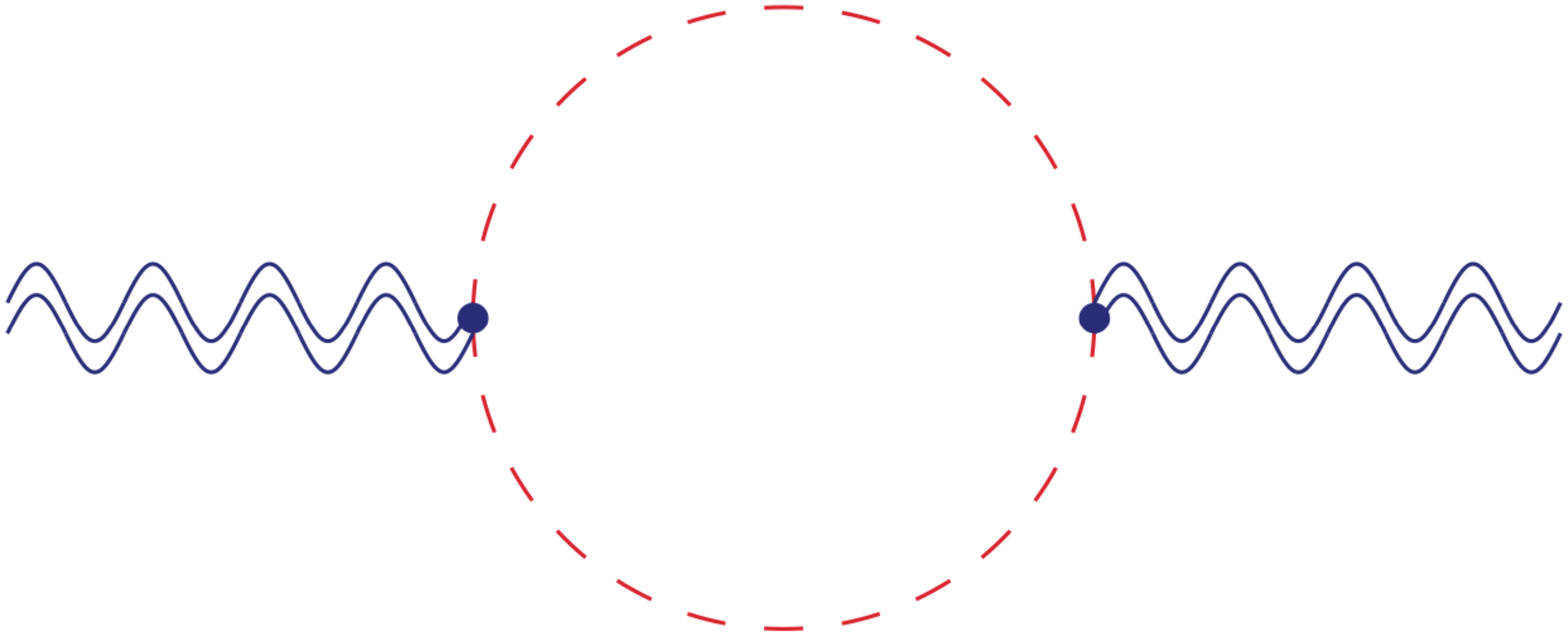}   
 \vspace{-1cm}
 \end{center}
   \caption{\small The leading loop diagram that generates the $R^2$ term in the effective action. The dashed lines correspond to the Higgs field, while the external double lines represent gravitons.}
\label{HiggsContrToGraviton}
\end{figure}

\subsection{Higgs-to-Starobinsky inflation}\label{Starobinsky}

\vspace{0.1cm}

Let us start this section by commenting on fine-tunings in the couplings, a relevant issue as inflation is motivated by cosmological fine-tuning problems. The first equation in (\ref{beta-m}) has an important implication; the Feynman diagram that leads to this contribution is given in Fig. \ref{HiggsContrToGraviton}.  Generically Higgs inflation requires a rather large value of $\xi$, which implies  a strong naturalness bound   
\be |\alpha|\gtrsim \frac{\xi^2}{8\pi^2}.\label{natural-alpha0}\ee
A large value of $\xi$ is necessary at the classical level (see Eq. (\ref{xi large}) and the corresponding discussion). At quantum level one can obtain smaller values, but still $\xi\gg1$ \cite{Bezrukov:2009db,Salvio-inf}. 

 A possible exception is Higgs inflation at the critical point \cite{Bezrukov:2014bra}; however, $\xi \gtrsim10$ to fulfill the most recent observational bounds, $r\lesssim0.1$ \cite{Ade:2015tva}. Moreover, in previous analysis of Higgs inflation at the critical point the wave function renormalization of the Higgs field has been neglected, an approximation that is under control when $\xi$ is large \cite{Bezrukov:2009db}. 

 Since $\xi\gg 1$ generically, (\ref{natural-alpha0}) indicates that an additional
 $R^2$ term with such a large coefficient may participate in inflation. Therefore, we write the  following effective action:
\begin{equation} \Gamma= \int d^4x\sqrt{-g}\left[\Lag_{\rm SM}^{\rm eff}-\left(\frac{\bp^2 }{2}+\xi |\mathcal{H}|^2\right)R+\alpha  R^2\right], \label{Jordan-frame-total-quantum}\end{equation}
where the $\Lag_{\rm SM}^{\rm eff}$ part   corresponds to the effective SM action.
The scalar-tensor effective action is 
\begin{equation} \Gamma_{\rm st} = \int d^4x\sqrt{-g}\left[\frac12 (\partial \phi)^2-V_{\rm eff}-\frac12 \left(\bp^2 +\xi \phi^2\right)R+\alpha  R^2\right]. \nonumber\end{equation}
Here we have neglected the wave function renormalization of the Higgs because $\xi$ is large and we have fixed the unitary gauge. Moreover, $V_{\rm eff}$ is the SM effective potential.

As well-known, the $R^2$ term  corresponds to an additional scalar. In order to see this one can add to the action  the  term  $$ -  \int d^4x\sqrt{-g}\,\, \alpha\left(R+ \frac{\omega}{4\alpha}\right)^2,$$ where $\omega$ is an auxiliary field: indeed by using the $\omega$ field equation one obtains immediately that this term vanishes. On the other hand, after adding that term
\begin{equation} \Gamma_{\rm st} = \int d^4x\sqrt{-g}\left[\frac12 (\partial \phi)^2-V-\frac{f}{2}R -\frac{\omega^2}{16\alpha}\right],  \end{equation}
where $f\equiv \bp^2+\omega+\xi \phi^2$.

Note that we have the non-canonical gravitational term $-fR/2$. Like we did in section \ref{model},  we can go to the Einstein frame (where we have instead  the canonical Einstein term $-\bar M_{\rm Pl}^2 R_E/2$)
by performing a conformal transformation,
\be g_{\mu\nu}\rightarrow  \frac{\bar M_{\rm Pl}^2}{f}g_{\mu\nu} .\ee
One obtains \cite{Kannike:2015apa}
\begin{equation} \Gamma_{\rm st} = \int d^4x\sqrt{-g}\left[\Lag_{\rm \phi z}-U_{\rm eff}-\frac{\bp^2}{2}R \right], \label{Gammast}\end{equation}
where 
\be  \Lag_{\rm \phi z} \equiv \frac{6\bp^2}{z^2}
 \frac{(\partial \phi)^2 + (\partial z)^2}{2},\nonumber \ee
 \be U_{\rm eff}(\phi,z)\equiv  \frac{36\bar M_{\rm Pl}^4}{z^4}\bigg[{V_{\rm eff}(\phi)}+   \frac{1}{16\alpha}\left(\frac{z^2}{6} -\bp^2- \xi  \phi^2\right)^2 \bigg]\nonumber\ee
 and we have introduced the new scalar $z=\sqrt{6f}$.

Notice that when $\alpha\rightarrow 0$, the potential $U_{\rm eff}$ forces $z^2=6(\bp^2 +\xi \phi^2)$ and we recover the Higgs inflation action. For large $\alpha$  (as dictated by a large $\xi$), this conclusion cannot be reached. 
The absence of runaway directions in $U_{\rm eff}$ requires $\alpha>0$ and $\lambda> 0$, which is possible within the pure SM (without gravity \cite{Loebbert:2015eea}), although in tension\footnote{Such tension, however, can be  be eliminated by adding to the SM well-motivated new physics, which solve its observational problems \cite{Salvio:2015cja}.} with the measured values of some electroweak observables \cite{Buttazzo:2013uya,Salvio-inf}.  
Ref. \cite{Kannike:2015apa} studied a system that includes (\ref{Gammast}) as a particular case\footnote{Ref. \cite{Kannike:2015apa} has an additional scalar which, however, can be consistently decoupled by taking its mass large enough. For another treatment of the dynamical system in (\ref{Gammast}) see Ref. \cite{Torabian:2014nva}.}. It was found that inflation is never dominated by the Higgs, because its quartic  self-coupling $\lambda$
(which we assume to be positive for the argument above) is unavoidably larger than the other scalar couplings, taking into account its RG flow.
Even assuming that the Higgs has a dominant initial  value,
in our two-field context inflation starts only after the field evolution has reached an attractor where $\phi$ is subdominant. We have checked that this happens also  when $\xi$ is large. 

Therefore, the predictions are closer to those of Starobinsky inflation, which are distinct from the Higgs inflation ones \cite{Bezrukov:2011gp}.

\section{Conclusions}

In conclusion, we have studied two different aspects of standard Higgs inflation -  to seek how fine-tuned the initial conditions should be to fall 
into a slow-roll attractor solution in an approximate exponentially flat Higgs potential in the Einstein frame. We started with a large kinetic energy, and we  found that for an  initial kinetic energy density of order $10^{-3} \bp^4$ (this is the maximum allowed order of magnitude  to avoid quantum gravity corrections)  the inflaton VEV should be $\sim10\bp$ to sustain inflation long enough to give rise to enough e-folds.

In the second half of the paper, we focused on the question of viability of Higgs inflation in presence of large $\xi$, typically required for
explaining the observed CMB power spectrum and the right tilt. We found that   one would incur quantum corrections (at the lowest order) to the Ricci scalar, i.e. quadratic in Ricci scalar,  $\alpha R^2$, with a universality bound on $\alpha$ given by Eq.~(\ref{natural-alpha0}), unless  the initial value of $\alpha$ is fine-tuned. 
If one includes this $R^2$  term in the effective action, both the Higgs and a new scalar degree of freedom
are present. By taking  $\xi\sim 10^{2}-10^{4}$ and using the bound in Eq.~(\ref{natural-alpha0}), the potential would be effectively determined by the Starobinsky scalar component $z$, and the CMB predictions would be different from that of Higgs inflation.


\section*{Acknowledgments} \noindent We thank  Mikhail Shaposhnikov for valuable correspondence and Alexei A. Starobinsky for discussing.  AM is supported by the STFC grant ST/J000418/1. AM also acknowledges the kind hospitality from IPPP, Durham, during the course of this work.
AS is supported by the Spanish Ministry of Economy and Competitiveness under grant FPA2012-32828, 
Consolider-CPAN (CSD2007-00042), the grant  SEV-2012-0249 of the ``Centro de Excelencia Severo Ochoa'' Programme and the grant  HEPHACOS-S2009/ESP1473 from the C.A. de Madrid.









\end{document}